\documentclass[reprint,
 amsmath,amssymb,aps,prx,footinbib]{revtex4-2}
\usepackage{times}
\usepackage{graphicx}
\usepackage{dcolumn}
\usepackage{bm}
\usepackage{mathtools}
\usepackage[english]{babel}
\usepackage{physics}
\usepackage{enumitem}
\usepackage{mathrsfs}
\usepackage{xfrac}
\usepackage{xr-hyper}
\usepackage{hyperref}
\usepackage{footmisc}
\usepackage{hypcap}
\usepackage{makecell}
\usepackage{url}
\usepackage{ifsym}
\usepackage{bbold}
\usepackage{gensymb}
\usepackage{multirow}
\usepackage{array}
\usepackage{longtable}
\usepackage{siunitx}
\usepackage{nicefrac}
\usepackage[dvipsnames]{xcolor}
\usepackage{ulem}
\usepackage{comment}
\usepackage{amsmath}

\newcommand{\ci}{\mathrm{i}}
\newcommand{\ee}{\mathrm{e}}
\newcommand{\lr}[1]{\left(#1\right)}
\newcommand{\lrsq}[1]{\left[#1\right]}
\newcommand{\tp}{\mathrm{p}}

\newcommand{\vna}{\boldsymbol{\nabla}}

\AtBeginDocument{\RenewCommandCopy\qty\SI}

\begin{document}

\title{Polariton Fluids as Quantum Field Theory Simulators on Tailored Curved Spacetimes}

\author{K\'evin Falque$^1$}
\author{Adri\`a Delhom$^2$}
\author{Quentin Glorieux$^1$}
\author{Elisabeth Giacobino$^1$}
\author{Alberto Bramati$^{1\dagger}$}
\author{Maxime J. Jacquet$^{1\dagger}$}

\affiliation{
$^1$ Laboratoire Kastler Brossel, Sorbonne Universit\'{e}, CNRS, ENS-Universit\'{e} PSL, Coll\`{e}ge de France, Paris 75005, France\\
$^2$ Department of Physics and Astronomy, Louisiana State University, Baton Rouge, LA 70803, U.S.A\\
$^\dagger$ correspondence to maxime.jacquet@lkb.upmc.fr and alberto.bramati@lkb.upmc.fr}

\begin{abstract}%
Quantum fields in curved spacetime exhibit a wealth of effects like Hawking radiation from black holes.
While quantum field theory in black holes can only be studied theoretically, it can be tested in controlled laboratory experiments.
In experiments, a fluid going from sub- to supersonic speed creates an effectively curved spacetime for the acoustic field, with a horizon where the speed of the fluid equals the speed of sound.
The challenge to test predictions like the Hawking effect in such systems lies in the control of the spacetime curvature and access to the field spectrum thereon.
Here, we create tailored stationary effective curved spacetimes in a polaritonic quantum fluid of light in which either massless or massive excitations can be created, with smooth and steep horizons and various supersonic fluid speeds.
Using a recently developed spectroscopy method we measure the spectrum of collective excitations on these spacetimes, crucially observing negative energy modes in the supersonic regions, which signals the formation of a horizon.
Control over the horizon curvature and access to the spectrum on either side demonstrates the potential of quantum fluids of light for the study of field theories on curved spacetimes, and we discuss the possibility of investigating emission and spectral instabilities with a horizon or in an effective Exotic Compact Object configuration.
\end{abstract}
\maketitle

Quantum field theory (QFT) in curved spacetimes predicts the amplification of field excitations and the occurrence of  classical and quantum correlations, as in the Hawking effect for example.
This raises the interest for experiments in which the curvature of spacetime can be controlled and correlations measured to test QFT predictions~\cite{almeida_analogue_2023,jacquet_next_2020,barcelo_analogue_2011}.
This is typically done with fluids accelerating from sub- to supersonic speeds: acoustic excitations are dragged by the supersonic flow, effectively trapped inside an acoustic horizon~\cite{visser_acoustic_1998}.
Quantum fluctuations of the acoustic field are predicted to yield entangled emission across the horizon~\cite{Unruh_experimental_1981}.

Here we use a polaritonic fluid of light~\cite{jacquet_polariton_2020} whose velocity profile we control to tailor the curvature of spacetime at the horizon and measure the field spectrum on either side of it.

One-dimensional effectively curved spacetimes realized in a number of experimental systems have enabled the observation of the Hawking effect from classical input states ~\cite{philbin_fiber-optical_2008,weinfurtner_measurement_2011,euve_observation_2016,munoz_de_nova_observation_2019,drori_observation_2019,shi_quantum_2023,tamura_2Dhorizon_2024}.
Meanwhile, two-dimensional spacetimes with rotation~\cite{torres_rotational_2017,torres_quasinormal_2020,patrick_backreaction_2021,braidotti_measurement_2022,svancara_rotating_2024} as well as time-dependent spacetimes~\cite{jaskula_acoustic_2012,clark_collective_2017,eckel_rapidly_2018,viermann_quantum_2022,tajik_experimental_2023,zenesini_false_2024} have also been implemented.

Most fluid systems consider the acoustic field $\phi_1(t,x,y)$ (small fluctuations around the mean-field) whose wave equation in an inviscid, barotropic, irrotational and conservative fluid $\phi_0(t,x,y)$ can be written in  covariant notation as $\frac{1}{\sqrt{|\eta|}}\partial_\mu\left(\sqrt{|\eta|}\eta^{\mu\nu}\partial_\nu\phi_1\right)=0$ in terms of the metric tensor
\begin{equation}
    \label{eq:metric}
    \eta_{\mu\nu}\propto\begin{pmatrix}
-\left({c_\mathrm{s}}^2-{v_0}^2\right) & -v_0^x & -v_0^y\\
-v_0^x & 1 & 0\\
-v_0^y & 0 & 1
\end{pmatrix}
\end{equation}
of determinant $\eta$.
The metric tensor components are controlled by the relative motion of $\phi_1$ (at the speed of sound ${c_\mathrm{s}}$) and $\phi_0$ (at velocity $\pmb{v_0}=v_0^x \pmb{e_x}+v_0^y\pmb{e_y}$).
For example, if a stationary flow of $\phi_0$ spatially changes (along $x$ only) such that its velocity exceeds the speed of sound at $x_\mathrm{H}$, excitations of $\phi_1$ will be trapped inside the supersonic region beyond the so-formed acoustic horizon at $v_0(x_\mathrm{H})=c_\mathrm{s}(x_\mathrm{H})$.
The spatial profile of $\phi_0$ defines the local causal structure seen by $\phi_1$, with the fluid velocity $\pmb{v_0}$ and speed of sound $c_\mathrm{s}$ profiles setting the effective curvature of the spacetime~\cite{visser_acoustic_1998}, inhomogeneities playing the role of tidal forces.
Horizons in fluid experiments act as static potentials on which excitations of $\phi_1$ scatter, leading to the mixing of creation and annihilation operators (i.e. the squeezing of field excitations)~\cite{agullo_event_2022}.
As in black holes, this results in an energetic instability of the (effective) spacetime due to the emission of entangled quanta~\cite{parker_particle_1968,hawking_black_1974,hawking_particle_1975,wald_particle_1975,Unruh_experimental_1981,agullo_entanglement_2023}.

Measurement of quanta and their correlations would enable  the dynamics of entanglement in various configurations~\cite{parentani_entanglementhorizon_2010,agullo_robustness_2022} to be tested experimentally, e.g. where quasi-normal modes could be excited and/or where an ergosurface surrounds the horizon~\cite{jacquet_quantum_2023,agullo_entanglement_2023}.
In these situations, extra modes of $\phi_1$ besides those usually involved in the Hawking effect interact with the horizon and modify the emission.
In this scheme, the properties of horizon emission itself have to be well controlled and understood.
Specifically, smooth horizons are predicted to yield a quasi-thermal spectrum, as in astrophysical black holes~\cite{fabbri_steplike_2011}.
To the contrary, when the horizon is steep the spectrum is dominated by the frequency-dispersion of $\phi_1$, which is nonlinear beyond the low wavenumber regime in all experimental systems~\cite{macher_black/white_2009,Recati_2009,finazzi_dispersion_2012}.
The key is to control the curvature of the effective spacetime, i.e. to create smooth or steep transitions between the sub- and supersonic regions.

While the Hawking effect is robust against departures from the relativistic setting in experimental systems~\cite{unruh_sonic_1995,brout_hawking_1995,corley_hawking_1996}, the resulting dynamics of entanglement are ruled by the specific shape of the dispersion~\cite{michel_phonon_2016,isoard_departing_2020,jacquet_influence_2020}, which determines the field theory under consideration.
Notably, we show here that, beyond phononic excitations (linear low wavenumber dispersion) as in other systems, transsonic polaritonic fluids can also support excitations with a massive, relativistic dispersion.
Consequently, the impact of mass on non-perturbative QFT predictions can be assessed.

\begin{figure*}[ht]
    \centering
    \includegraphics[width=\textwidth]{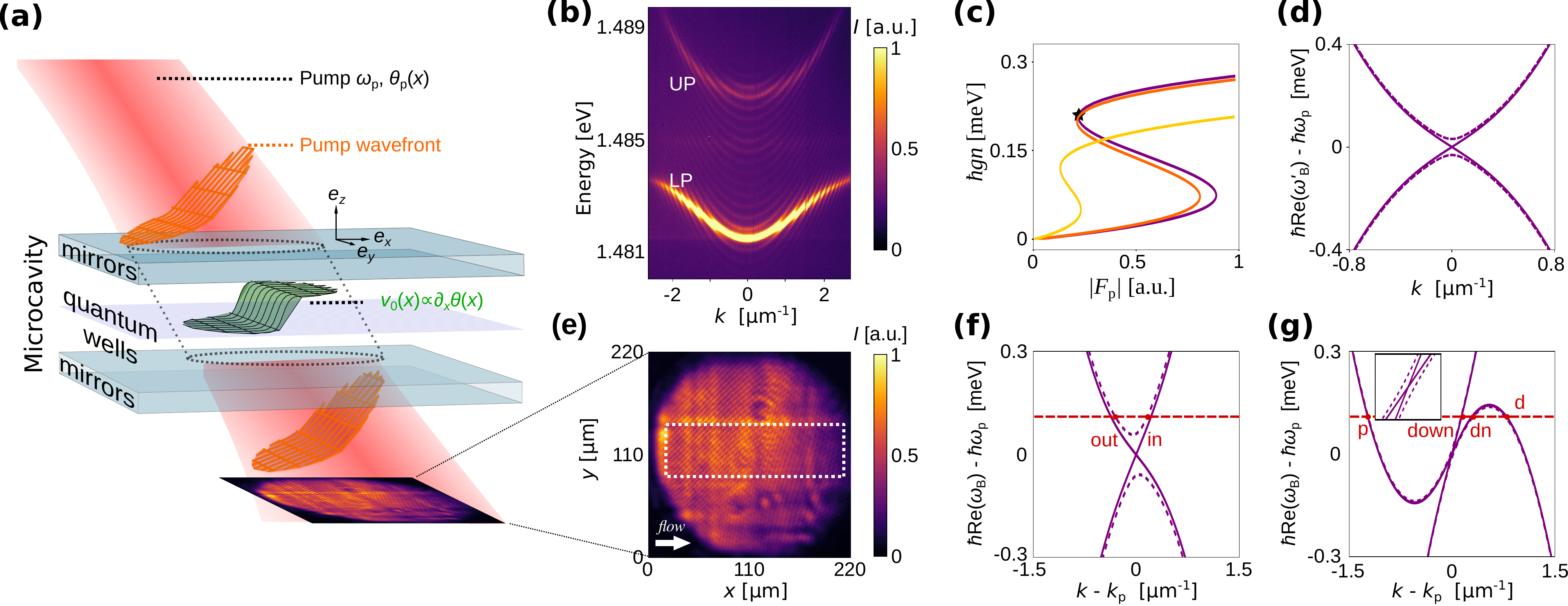}
    \caption{\textbf{Polaritonic fluid of light}. In this figure, $\delta(0)= \SI{56}{\giga\hertz} \rightarrow \hbar\delta(0)=\SI{0.2}{\milli\electronvolt}$.
    \textbf{(a)} \textbf{Spatial profile and wavefront of the pump beam} before (orange) and after (orange) the microcavity.
    The spatial derivative of the phase gives the velocity profile of the fluid in the cavity plane (green).
    \textbf{(b)} \textbf{Polariton eigenstates} in the strong coupling regime. UP, upper polaritons; LP, lower polaritons.
    Modulations appear at a frequency corresponding to the free spectral range of the sample substrate. These do not affect polariton physics.
    \textbf{(c)} \textbf{Optical bistability of the polariton density}. Pump incident on the cavity at different $k_\mathrm{p}$ (analytical calculation): purple, $k_\mathrm{p}=\SI{0}{\per\micro\meter}$ (the black star marks the point for which $gn=\delta(0)$); orange, $k_\mathrm{p}=\SI{0.11}{\per\micro\meter}$; yellow, $k_\mathrm{p}=\SI{0.55}{\per\micro\meter}$.
    \textbf{(d)} \textbf{Dispersion relation of collective excitations in the fluid rest-frame} Eq.~\eqref{eq:ffbogo} (analytical calculation): continuous curve, $\abs{F_\mathrm{p}(gn=\delta)}$ (black star in \textbf{(c)}); dashed curve, $\abs{F_\mathrm{p}(gn>\delta)}$. 
    \textbf{(e)} \textbf{Intensity of the light after the cavity}, showing the mean-field density of polaritons in the cavity.
    $45\degree$ fringes are artifacts due to a filter in the CMOS camera.
    Measurements in subsequent figures are averaged over the width of the white-dashed rectangle.
    \textbf{(f)}-\textbf{(g)} \textbf{Dispersion relation of collective excitations in the laboratory frame} Eq.~\eqref{eq:lfbogo} with sub- and supercritical flow velocities (analytical calculation): continuous curve, $\abs{F_\mathrm{p}|_{gn=\delta}}$ (black star in \textbf{(c)}); dashed curve, $\abs{F_\mathrm{p}gn>\delta)}$.
    \textbf{(f)} $k_\mathrm{p}=\SI{0.11}{\per\micro\meter}$.
    \textbf{(g)} $k_\mathrm{p}=\SI{0.55}{\per\micro\meter}$.
    The inset shows the behavior of the curves near $k-k_\mathrm{p}=0$.
    Red-dashed line, example frequency at which the Hawking effect occurs, mixing ingoing modes \textit{in}, \textit{p} and \textit{d}, and giving outgoing modes \textit{out} (Hawking radiation), \textit{down} and \textit{dn} (the partner).}
    \label{fig:fig1}
\end{figure*}

In our experiment we engineer tailored spacetimes that enable the investigation of this rich phenomenology via the precise measurement of $\phi_1$'s properties.
We demonstrate all-optical control of the curvature of the effective spacetime, as well as direct access to the field theory at play.
We engineer  smooth and steep horizons, and, exploiting the driven-dissipative nature of the polaritons, we use a recently developed coherent probe spectroscopy method~\cite{claude_high-resolution_2022,claude_spectrum_2023} to measure the spectrum of collective excitations $\phi_1$.

\section{Polariton fluid and collective excitations}
When coherent light is shone upon quantum wells embedded in an optical microcavity [Fig~\ref{fig:fig1} \textbf{(a)}], the optical field can strongly couple to the bound electron-hole excitations (excitons) of the material.
The Hamiltonian describing the excitonic and electromagnetic fields then has two new eigenstates called the upper (UP) and lower (LP) polariton [Fig~\ref{fig:fig1} \textbf{(b)}].
Due to their hybrid light-matter nature, these bosonic quasi-particles inherit both the low effective mass of the cavity photons and the nonlinear exchange interactions of the excitons.
Consequently, they collectively behave as a fluid of light~\cite{carusottoQuantumFluidsLight2013a}.
In our experiment, only the lower polaritons are excited.
Their mean-field dynamics are described by a driven dissipative Gross-Pitaevskii equation (GPE)
\begin{equation}\label{eq:ddGPE}
    \begin{aligned}
    \ci\hbar\partial_t \Phi(\pmb{r},t)  = & \left(\hbar\omega_0 - \frac{\hbar^2}{2m^*} \nabla^2 + \hbar g\lvert\Phi(\pmb{r},t)\rvert^2
    - \ci\frac{\hbar\gamma}{2}\right) \Phi(\pmb{r},t)\\&+F_\mathrm{p}(\pmb{r},t)  ,       
    \end{aligned}
\end{equation}
where $g$ is the repulsive ($g>0$) polariton-polariton interaction, $\omega_0$ is the polariton angular frequency at zero wavenumber $k=0$ and $m^*$ is the effective polariton mass.
The pump term $F_\mathrm{p}(\pmb{r},t)$ describes the electromagnetic field of the laser that injects photons into the cavity.
These photons bind with the excitons to form polaritons which eventually decay into a photon exiting the cavity with rate $\gamma/2$.
To reach a steady-state, a continuous wave (CW) laser is used to continuously compensate for the losses.

Here, the polaritons are pumped quasi-resonantly, so the phase of the laser pump sets that of the polariton field $\Phi$: $\theta(\pmb{r})=\theta_\mathrm{p}(\pmb{r})$.
The steady state of Eq~\eqref{eq:ddGPE} is such that the polariton density $n(\pmb{r},t)=\lvert\Phi(\pmb{r},t)\rvert^2$ (measured via the output intensity $I(\pmb{r},t)$) is related to the incident intensity $\lvert F_\mathrm{p}\rvert^2$ via a nonlinear relation depending on the effective detuning $\delta({k_\mathrm{p}})=\omega_\mathrm{p}-\omega_0-\frac{\hbar k_\mathrm{p}^2}{2m^*}$  between the frequency of the driving laser (pump) and
the polariton frequency at the pump wavevector $\pmb{k_\mathrm{p}}$.
Importantly, when $\delta({k_\mathrm{p}})>\sqrt{3}\gamma/2$, the system enters a bistable regime~\cite{baas_bista_2004} [Fig~\ref{fig:fig1} \textbf{(c)}].

Collective excitations $\phi_1$ of the polariton fluid can be studied by linearising the GPE around a steady-state via $\Phi(\pmb{r}, t) = \left(\phi_0 + \phi_1\right)\ee^{\ci\left(\theta_{\tp}(\pmb{r})-\omega_{\tp} t\right)}$.
In a spatially homogeneous region, their spectrum is given by the Bogoliubov dispersion relation~\cite{carusottoQuantumFluidsLight2013a}, which in the fluid rest-frame may be written as
\begin{equation}
    \label{eq:ffbogo}
\omega_\mathrm{B}^{\prime\pm}({\delta k})=-\frac{\ci\gamma}{2} \pm \sqrt{\left(\frac{\hbar {\delta k}^2}{2m^*}-\delta({k_\mathrm{p}}) +2gn +g_\mathrm{r} n_\mathrm{r}  \right)^2-(gn)^2},
\end{equation}
where $\delta k=k-k_\mathrm{p}$ while $g_\mathrm{r} n_\mathrm{r}=\beta\times gn$ ($\beta=cst\geq0$) accounts for possible modifications to the interaction energy under the effect of a long-lived exciton reservoir not coupled to the light field~\cite{claude_spectrum_2023} [Appendix~\ref{App:cs}].

Solutions $\omega_\mathrm{B}^{\prime+}$ ($\omega_\mathrm{B}^{\prime-}$) have higher (lower) fluid rest-frame frequency than that of the pump $\omega_\mathrm{p}$.
They are often referred to as the normal and ghost branch, respectively.

Fig~\ref{fig:fig1} \textbf{(d)} shows the real part of the dispersion relation~\eqref{eq:ffbogo} for two different values of pump intensity $\abs{F_\mathrm{p}}$.
When the pump intensity is such that $gn=\delta-g_\mathrm{r} n_\mathrm{r}$ (solid curve), the dispersion is linear (phononic) at low $k$, giving the usual form of the speed of sound, $c_\mathrm{s}=\sqrt{\hbar gn/m^*}$~\cite{claude_spectrum_2023}.
Remarkably, when the pump intensity is such that $gn>\delta-g_\mathrm{r} n_\mathrm{r}$ (dashed curve), the dispersion becomes parabolic and the sound cones are then given by $c_\mathrm{B}=\sqrt{\hbar(2gn_0-\delta(k_\mathrm{p})+g_\mathrm{r}n_\mathrm{r})/m^*}$ [Appendix~\ref{App:metric}].
The field $\phi_1$ acquires an effective mass $m_{\mathrm{det}}$ that is experimentally controlled by the pump intensity and effective detuning $\delta(k_\mathrm{p})$.
Control over the mass of $\phi_1$ is generally absent in other quantum simulator platforms such as atomic BECs.

\begin{figure*}[ht]
    \centering
    \includegraphics[width=\textwidth]{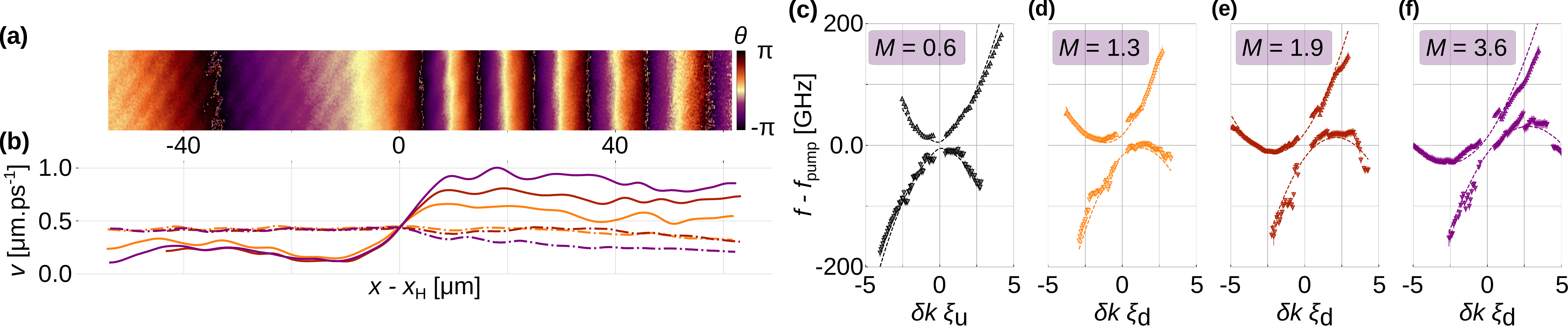}
    \caption{\textbf{Smooth horizon}.
    \textbf{(a)} Typical measured phase of the fluid $\theta(x)$ inside the white rectangle of Fig.~\ref{fig:fig1}.
    \textbf{(b)} Fluid velocity profiles.
    Solid lines, $v_0(x)$; dot-dashed lines, $c_\mathrm{s}(x)$. Orange, $M_\mathrm{u}=0.6$ and $M_\mathrm{d}=1.4$; red, $M_\mathrm{u}=0.4$ and $M_\mathrm{d}=1.9$; purple $M_\mathrm{u}=0.5$ and $M_\mathrm{d}=3.6$.
    \textbf{Excitation spectra} Eq.~\eqref{eq:lfbogo}: \textbf{(c)} typical upstream region; \textbf{(d)}-\textbf{(f)} downstream region.
    Up-triangles, positive-norm branch $\omega_\mathrm{B}^+$; down-triangles, negative-norm branch $\omega_\mathrm{B}^-$; dashed lines, fit with free parameter $gn$. Error bars (mostly of same size as the triangles) represent the precision of wavelength measurements with the lambdameter.
    \label{fig:ModeStructure}}
\end{figure*}

\section{Field theory and effectively curved spacetime}
As usual, the first-order system for $\phi_1$ and its conjugate, obtained by linearising the GPE~\eqref{eq:ddGPE}, can be recast into a fourth-order partial differential equation for $\phi_1$ that becomes a second order relativistic equation in the hydrodynamic regime~\cite{visser_acoustic_1998}.
In the case in which $\phi_1$ has mass $m_\mathrm{det}$ (neglecting terms related to dissipation, see Appendix~\ref{App:metric} for details), this equation takes the form of a massive Klein-Gordon equation on an effective curved spacetime
\begin{equation}
\label{eq:massmetric}
    \left[\frac{1}{\sqrt{|\eta|}}\partial_\mu\sqrt{|\eta|}\eta^{\mu\nu}\partial_\nu-\frac{(m_\mathrm{det})^2}{\hbar^2}\right]\phi=0,
\end{equation}
where $\eta_{\mu\nu}$ is the acoustic metric~\eqref{eq:metric}.

Eq~\eqref{eq:massmetric} has a conserved quantity called the symplectic product, which, for two solutions $\phi_{\mathrm{\alpha},\mathrm{\beta}}$, is defined by 
\begin{equation}\label{eq:SP}
    (\phi_\mathrm{\alpha},\phi_\mathrm{\beta}) \coloneqq \ci \int  \left(\phi_\mathrm{\alpha}^*\,  \pi_\mathrm{\beta}-\pi_\mathrm{\alpha}^*\, \phi_\mathrm{\beta}\right)\mathrm{d}^2\pmb{r}\,,    
\end{equation}
with $\pi_{\alpha,\beta}$ the canonical conjugate momentum of $\phi_{\alpha,\beta}$ (in the fluid rest-frame $\pi_{\alpha,\beta}=\sqrt{|\eta|}\partial_t\phi_{\alpha,\beta}$).
This defines the conserved symplectic (pseudo-)norm of a solution as $Q_{\phi}\coloneqq (\phi,\phi)$.
The symplectic norm is not positive definite and, in our case, $\omega_B^{\prime+}$ ($\omega_B^{\prime-}$) modes have positive (negative) norm~\cite{macher_black/white_2009}\footnote{The sign of the norm of the solutions corresponding to positive and negative branches of the linear dispersion can be computed from the conserved symplectic product. Its conservation holds also beyond hydroynamics, i.e. when nonlinearities of the dispersion are taken into account.}.

Consider now a spatially inhomogeneous configuration in which the polariton fluid accelerates along $x$ only, i.e. the phase gradient along $y$ is zero and $\pmb{k_\mathrm{p}}=k_\mathrm{p}(x)\pmb{e_\mathrm{x}}$.
Using the Madelung representation  $\phi_0=\sqrt{n(x)}e^{i\theta(x)}$, we consider the so-called ``waterfall'' velocity profile~\cite{Leboeuf_waterfall_2003}
\begin{align}
    \label{eq:velocityprofile}
    v_0(x) &= \frac{\hbar}{m^*}\partial_x \theta(x)\nonumber\\
         &= \frac{v_\mathrm{d}-v_\mathrm{u}}{2}\,tanh\left(\frac{x-x_\mathrm{H}}{w_\mathrm{H}}\right)+\frac{v_\mathrm{d}+v_\mathrm{u}}{2}.
\end{align}
The fluid has two asymptotic regions: upstream, where the fluid velocity $v_0(x\ll x_\mathrm{H})=v_\mathrm{u}$, and downstream, where $v_0(x\gg x_\mathrm{H})=v_\mathrm{d}$.
These homogeneous regions are separated by a transition of width $w_\mathrm{H}$, centered at $x=x_\mathrm{H}$, where the fluid velocity increases from $v_\mathrm{u}$ to $v_\mathrm{d}$.
Likewise, the density $gn(x)$ is homogeneous in each asymptotic region.
Acceleration drags the $\omega_\mathrm{B}^{\prime\pm}$ modes along $x$, resulting in a Doppler shift of the frequency~\eqref{eq:ffbogo} when measured in the laboratory frame
\begin{align}
    \label{eq:lfbogo}
\omega_\mathrm{B}^{\pm}&= -\frac{\ci\gamma}{2}+ v_0(x) \delta k(x)\,\pm\\ &\sqrt{\left(\frac{\hbar \delta k(x)^2}{2m^*}-\delta(k_\mathrm{p}) +2gn(x) +g_\mathrm{r} n_\mathrm{r}    \right)^2-(gn(x))^2},\nonumber
\end{align}
with $\delta k(x)=k-k_\mathrm{p}(x)$.

As a function of the Doppler shift $v_0(x) \delta k(x)$, the branches of the dispersion~\eqref{eq:lfbogo} assume different shapes.
For subcritical flows, the branches are only slightly deformed [Fig.~\ref{fig:fig1} \textbf{(f)}].
For supercritical flows, the branches are deformed such that the negative-norm branch is pulled to positive laboratory-frame frequencies (or, conversely, the positive-norm branch is pulled to negative laboratory-frame frequencies) [Fig.~\ref{fig:fig1} \textbf{(g)}].
Negative-norm waves at positive frequencies and positive-norm waves at negative frequencies are negative energy waves ($\mathrm{sign}(E)=\mathrm{sign}(\omega_\mathrm{B})\times \mathrm{sign}(Q_{\phi_1})$~\cite{Bogolyubov:1947zz}).

In conservative fluids, sub- and supercritical flows are discriminated by the Mach number $M\coloneqq v_0/c_\mathrm{s}=1$.
Instead, in our driven-dissipative fluid, the minimum velocity to have a supercritical dispersion is larger than $c_\mathrm{s}$ [Appendix~\ref{app:vctirical}].
As a result, the condition $M=1$, which defines the acoustic horizon in the hydrodynamic limit does not necessarily coincide with the condition to excite negative energy waves in the system.

If the fluid goes from sub- to supercritical flow velocity, positive and negative energy modes of $\phi_1$ can be simultaneously excited at a single, positive laboratory-frame frequency $\omega$ (red-dashed line in Fig.~\ref{fig:fig1} \textbf{(f)}-\textbf{(g)}).
In a scattering process between excitations coming towards the horizon from either side, norm-mixing will occur, giving rise to amplification by the Hawking effect (i.e. two-mode squeezing).

The strength of correlations across the horizon depends on both the horizon steepness through
\begin{equation}\label{eq:surfacegravity}
    \kappa \coloneqq \frac{1}{2c_\mathrm{s}(x)}\frac{d}{dx}[v^2_0(x)-c^2_\mathrm{s}(x)]|_{x_\mathrm{H}},
\end{equation}
the hydrodynamic equivalent to the surface gravity~\cite{barcelo_analogue_2011}, and the frequency detuning $\delta(k_\mathrm{p})$ in each asymptotic regions~\cite{jacquet_analogue_2022}.
The spectrum shows a clear signature of $m_\mathrm{det}$ [see opening of mass gap in Fig.~\ref{fig:fig1} \textbf{(d)}-\textbf{(g)} and remarks in Appendix~\ref{app:vctirical}], resulting in a minimum frequency for pair production.
Similarly, superluminality of the dispersion and a finite asymptotic velocity in the supercritical region lead to a maximum frequency at which the amplification occurs~\cite{jacquet_analogue_2022}.
These features are observed in the shape of the spectrum (Doppler shift), signaling the formation of a horizon and the possibility for the Hawking effect to occur (appearance of negative energy waves), making the spectrum a key observable to determine the QFT at play.

\section{Experimental implementation} 
We study a semiconductor microcavity consisting of three InGaAs quantum wells sandwiched between two highly reflecting planar GaAs-AlGaAs Bragg mirrors.
The quantum wells, separated by GaAs barriers, are located at the three antinodes of the cavity, which has a finesse on the order of 3000.
By fitting the measured spectrum of [Fig~\ref{fig:fig1} \textbf{(b)}] with the theoretical expression~\cite{carusottoQuantumFluidsLight2013a}, we extract the effective polariton  mass at the working point, $m^*=7.0 \times 10^{-35}$ kg.
The experiments are performed in an open-flow helium cryostat ($T\approx \SI{4}{\kelvin}$).

The polariton fluid is generated by a CW Ti:sapphire laser with a sub-MHz linewidth.
The Gaussian beam is reflected on a spatial light modulator (SLM) which imprints the target phase $\theta_\mathrm{p}(x)=\int v_0(x)dx$ that determines the fluid velocity $v_0(x)$ at each point.
The phase profile (orange wavefront in Fig.~\ref{fig:fig1} \textbf{(a)}) is imaged in the plane of the cavity with two focal-length matched telescopes ($2f-2f$ configuration).
The resulting fluid has a Gaussian transverse intensity profile with an $\approx\SI{120}{\micro\meter}$ waist.

The mean-field properties of the polariton fluid are fully characterized by imaging the outgoing optical field on a CMOS camera.
Intensity measurements provide direct access to the polariton density distribution and, once the interaction energy $\hbar gn$ and the detuning $\delta$ are determined at one point in the fluid, a map of the sound velocity $c_\mathrm{s}(x)$ can be reconstructed ($\hbar gn$ is obtained by taking the reservoir contribution as $\beta=1.84$~\cite{claude_spectrum_2023} and fitting the spectral data with Eq.~\eqref{eq:lfbogo}, see Appendix~\ref{App:cs} for details).
Simultaneously, the spatial phase of the outgoing light is measured by off-axis interferometry, allowing the velocity field to be determined ($v_0(x)=\partial_x\theta$)~\cite{phaseutils}.
We are thus able to exhaustively characterize the effective spacetime in our experiment. 

The dispersion~\eqref{eq:ffbogo} is measured by coherent probe spectroscopy~\cite{claude_high-resolution_2022} using a second CW Ti:sapphire laser (probe) of \SI{30}{\micro\meter} waist.
When it resonates with the cavity (whose nonlinearity is modified by the pump), the probe creates small perturbations in the fluid generated by the pump and the probe light is transmitted.
The probe intensity is set about two orders of magnitude below that of the pump and is intensity-modulated at \SI{5}{\mega\hertz}.
The probe wavenumber $k_\mathrm{pr}$ is varied from -\SI{2}{\per\micro\meter} to \SI{2}{\per\micro\meter} with a resolution $0.0189\pm \SI{0.0005}{\per\micro\meter}$ with a second SLM that controls its angle of incidence.
For each value of $k_\mathrm{pr}$, the probe frequency $\omega_\mathrm{pr}$ is continuously scanned over a range of \SI{400}{\giga\hertz} around $\omega_\mathrm{p}$.
A controllable and $k-$tunable pinhole placed in the reciprocal space of the microcavity output plane selects the transmitted light and directs it on a photodiode whose electronic signal is filtered around the modulation frequency with a spectrum analyzer (zero-span mode, resolution bandwidth \SI{300}{\hertz}).
We monitor the probe transmission and identify resonance peaks whose maxima give $\mathcal{R}e(\omega_\mathrm{B}^\pm$).

While the normal branch $\omega_\mathrm{B}^+$ can be excited directly when $\omega_\mathrm{pr}\simeq \omega_\mathrm{B}^+(k_\mathrm{pr})$, direct excitation of the ghost branch is less efficient.
Instead, in the experiment we take advantage of the strong nonlinearity of the system to excite $\omega_\mathrm{B}^-$ by four-wave mixing ($2\omega_\mathrm{p}(k_\mathrm{p})=\omega_\mathrm{B}^+(k_\mathrm{pr}) + \omega_\mathrm{B}^-(-k_\mathrm{pr})$)~\cite{claude_spectrum_2023}.

\section{All-optical control of the effective curvature}
\subsection{Smooth horizon configurations}
We first study a smooth transition between the sub- and supercritical flow regions.
We implement the target velocity profile \eqref{eq:velocityprofile} with a transition width $w_\mathrm{H}=\SI{20}{\micro\meter}$. 
We study three profiles, with three values for the up- and downstream flow velocities $v_\mathrm{u}$ and $v_\mathrm{d}$.
All the measurements are made with $\delta(0)=\SI{56}{\giga\hertz}\rightarrow \hbar\delta(0)= \SI{0.2}{\milli \electronvolt}$.
At this detuning, the healing length $\xi=\sqrt{\hbar/m^*(2gn_0-\delta+g_{\rm r}n_{\rm r})}$ is typically $\SI{3.4}{\micro\meter}$ upstream and $\SI{4.0}{\micro\meter}$ downstream.

Fig~\ref{fig:ModeStructure} \textbf{(b)} shows the measured fluid velocity $v_0(x)$ (solid lines) and the sound velocity $c_\mathrm{s}(x)$ (dot-dashed  lines).
Fast variations around the mean in $v_0(x)$ and $c_\mathrm{s}(x)$ are due to the inhomogeneities of the cavity that locally modify the effective detuning.
Given the speed of sound $c_\mathrm{s}(x)\approx\SI{0.40}{\micro\meter\per\pico\second}$, the critical Mach number is $M_\mathrm{critical}=1.34$ [Appendix~\ref{App:cs}].

We measure Mach numbers $M_\mathrm{u}=0.6$ (orange), $M_\mathrm{u}=0.4$ (red) and $M_\mathrm{u}=0.5$ (purple) and $M_\mathrm{d}=1.3<M_\mathrm{critical}$ (orange), $M_\mathrm{d}=1.9$ (red) and $M_\mathrm{d}=3.6$ (purple).

\begin{figure}[ht]
    \centering
\includegraphics[width=\columnwidth]{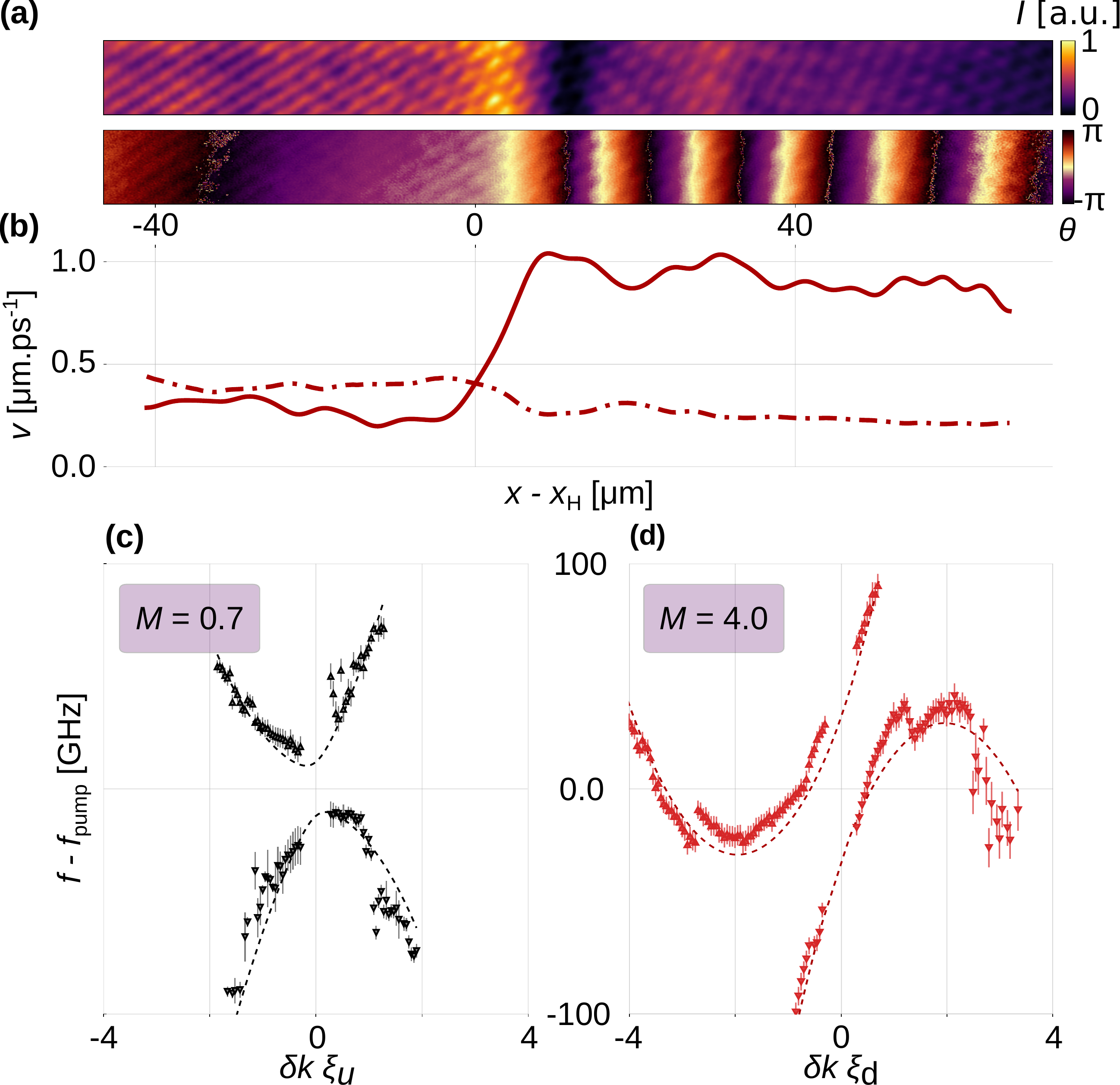}
    \caption{
    \textbf{Steep horizon}.    
    \textbf{(a)} Measured fluid density (top) and phase (bottom).
    \textbf{(b)} Measured fluid velocity profile.
    Solid line, $v_0(x)$; dashed line, $c_\mathrm{s}(x)$.
    \textbf{Excitation spectra} Eq.~\eqref{eq:lfbogo}: \textbf{(c)} upstream region. \textbf{(d)} downstream region; dashed lines, fit with free parameter $gn$. Error bars represent the precision of wavelength measurements with the lambdameter.}
    \label{fig:ModeStructure_steep}
\end{figure}

Fig~\ref{fig:ModeStructure} \textbf{(c)}-Fig~\ref{fig:ModeStructure} \textbf{(d)}-\textbf{(f)} show the excitation spectra~\eqref{eq:lfbogo} corresponding to the three spacetimes of \textbf{(b)}.
Note that, given the probe mode size in real space ($\SI{60}{\micro\meter}$), the apparatus resolution limits spectral measurements to $\abs{k}>\SI{0.10}{\per\micro\meter}$.
In all configurations, the upstream spectrum measured at $x<x_\mathrm{H}$ \textbf{(c)} has $\omega_\mathrm{B}^+(k) - v_0  \delta k_\mathrm{p}>0$ and  $\omega_\mathrm{B}^-(k) - v_0  \delta k_\mathrm{p}<0$.
Thus no negative energy waves are excited at $x<x_\mathrm{H}$, which confirms that the fluid velocity is subcritical in this region.

The three spectra taken at $x>x_\mathrm{H}$ in \textbf{(d)}-\textbf{(f)} all show a Doppler shift.
While in Fig~\ref{fig:ModeStructure} \textbf{(e)} and \textbf{(f)} the Doppler shift is large enough for $\omega_\mathrm{B}^-(k)$ to be pulled to positive laboratory-frame frequencies and $\omega_\mathrm{B}^+(k)$ to negative laboratory-frame frequencies, this is not the case in Fig~\ref{fig:ModeStructure} \textbf{(d)}.
In that case, $M_\mathrm{d}<M_\mathrm{critical}$ so the Doppler effect is not large enough to pull the positive and negative norm branch to positive and negative frequencies, respectively.
Therefore, no negative energy modes are excited, which means that no Hawking effect can occur.
 On the contrary, in the red and purple configurations, $M_\mathrm{d}$ is large enough for negative energy modes to be excited, meaning that the Hawking effect can occur in these configurations, for which we measure horizon steepness $\kappa=\SI{0.07}{\per\pico\second}$ (red) and $\kappa=\SI{0.08}{\per\pico\second}$ (purple).

\subsection{Steep horizon}
The effective detuning $\delta(k_\mathrm{p})$ reduces when $k_\mathrm{p}$ increases, thus lowering the values of $gn$ on the top branch of the bistability [Fig~\ref{fig:fig1} \textbf{(c)}] for large velocities.
We exploit this phenomenology to lower $c_\mathrm{s}(x>x_\mathrm{H})$ in the downstream region and increase the horizon steepness.

We implement the target profile~\eqref{eq:velocityprofile} with a large difference between $v_\mathrm{d}$ and $v_\mathrm{u}$, a transition width $w_\mathrm{H}=\SI{20}{\micro\meter}$ and a detuning $\delta(0)=\SI{71}{\giga\hertz} \rightarrow\hbar\delta(0)= \SI{0.3}{\milli\electronvolt}$.
Fig~\ref{fig:ModeStructure_steep} \textbf{(b)} shows $v_0(x)$ (solid lines) as well as $c_\mathrm{s}(x)$ (dot-dashed lines).
We measure $M_\mathrm{u}=0.7$ and $M_\mathrm{d}=4$, with $\kappa= \SI{0.11}{\per\pico\second}$.
Correspondingly, the up- and downstream spectra~\eqref{eq:lfbogo} \textbf{(c)}, \textbf{(d)} are sub- and supercritical.
$M_\mathrm{u,d}$ is not very different from the values in Fig.~\ref{fig:ModeStructure}.

We can control the horizon steepness $\kappa$ without changing the asymptotic properties of the spacetime, which allows to modify the strength of correlations~\cite{barcelo_analogue_2011,jacquet_analogue_2022} without changing the QFT at play.
This independent tuning of $\kappa$ and $M_\mathrm{u,d}$ is not possible in conservative quantum fluids such as atomic BECs, for example.
The fine control over the waterfall horizon geometry we demonstrate here is interesting to test recent tunneling models for the Hawking effect~\cite{delporro2024tunneling}.

\begin{figure}[ht]
    \centering
\includegraphics[width=\columnwidth]{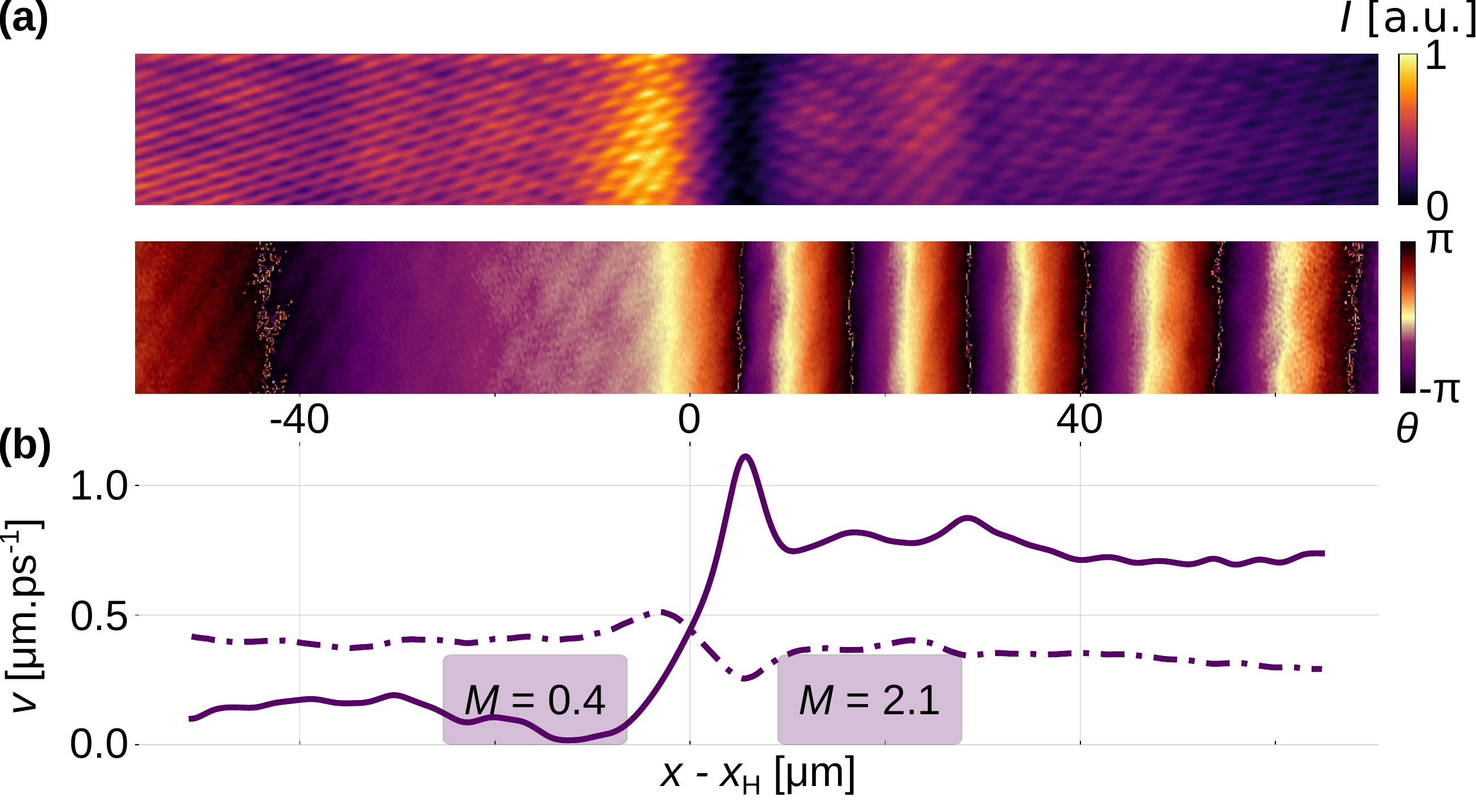}
    \caption{
    \textbf{Quasi-normal mode configuration}.  
    \textbf{(a)} Measured fluid density (top) and phase (bottom).
    \textbf{(b)} Measured fluid velocity profile. Solid line, $v_0(x)$: dot-dashed line, $c_\mathrm{s}(x)$.
    }
    \label{fig:qnm}
\end{figure}

\subsection{Quasi-normal mode configuration}
The profile~\eqref{eq:velocityprofile} can in fact allow for the creation of more complicated horizon configurations by imprinting a velocity peak immediately inside the horizon, as shown in Fig.~\ref{fig:qnm}.
We measure $M_\mathrm{u}=0.4$ and $M_\mathrm{u}=2.1$, so the spectrum of excitations in the asymptotic regions resembles those showed in Fig.~\ref{fig:ModeStructure} \textbf{(c)} and \textbf{(e)}, respectively.
The peak in $v_0$ and corresponding dip in $c_\mathrm{s}(x)$, have a half width at half maximum/minimum of \SI{3.3}{\micro\meter}.

As the dip is surrounded by two regions of higher density, the dip creates an effective resonator for excitations of $\phi_1$.
As in all resonators, this leads to the appearance of a metastable excitation mode of $\phi_1$ called a quasi-normal mode: here specifically, a negative-energy standing wave can establish itself inside the resonator and tunnel-couple with positive-energy propagating waves on either side~\cite{jacquet_quantum_2023}.
This quasi-normal mode is spontaneously excited by quantum vacuum fluctuations of $\phi_1$.
As a result, the Hawking spectrum peaks at the frequency of the quasi-normal mode, bearing signatures of the near-horizon geometry.

We have demonstrated control of the fluid density along the optical bistability (which is key to controlling $\phi_1$'s mass $m_\mathrm{det}$ and to observe  the Hawking effect in polariton fluids~\cite{jacquet_analogue_2022}), control over the horizon steepness and the realization of a configuration in which quasi-normal modes can be excited.
We have complete control on the curvature of the effective spacetime in the experiment.
This is advantageous over other methods relying e.g. on nanofabrication to shape the polariton flow~\cite{Nguyen}, or over atomic BEC experiments, showing that our platform is versatile and enables the investigation of QFT on tailored curved spacetimes.

\section{Discussion} Polaritonic fluids of light are an all-around platform for quantum simulation~\cite{boulierMicrocavityPolaritonsQuantum,jacquet_polariton_2020}.
In this experiment, we have shown how to spatially engineer a purely one-dimensional transcritical flow by all-optical control.
We have tapped-into the intrinsic non-equilibrium properties of the polariton fluid to create tailored sub- to supercritical transitions: we have implemented the canonical ``waterfall'' velocity profile~\cite{Leboeuf_waterfall_2003} and finely tuned the horizon steepness from $\kappa=\SI{0.07}{\per\pico\second}$ to \SI{0.11}{\per\pico\second}.
This is particularly interesting for the investigation of the Hawking effect whose flux (number of particles per unit time and unit bandwidth) increases with steepness and is given (in the hydrodynamic regime) by~\cite{balbinot_nonlocal_2008,larre_quantum_2012,michel_phonon_2016}
\begin{equation}
    \label{eq:Hflux}
    I_\omega^\mathrm{HR}=\frac{1}{e^{\omega/2\pi\kappa}-1}.
\end{equation}
For example, we take $f-f_{pump}=\SI{20}{\giga\hertz}$ and calculate an increase from $I_\omega^\mathrm{HR}=3$ for smooth horizons to $I_\omega^\mathrm{HR}=5$ for steep horizons.
The tunability of $\kappa$ in our system allows changing $I_\omega^\mathrm{HR}$ without changing the spectrum in the asymptotic regions, therefore increasing the total Hawking flux $\int I_\omega^\mathrm{HR} \mathrm{d}\omega$.  
For the smoothest horizon (whose full spectrum is quasi-thermal~\cite{fabbri_steplike_2011}) we obtain a Hawking `temperature' of $T^\mathrm{HR}=h\kappa/ k_{\rm B} =\SI{3}{\kelvin}$.
On the other hand, the spectrum of steep horizons is dominated by the nonlinear frequency dispersion~\cite{michel_phonon_2016,isoard_departing_2020,jacquet_influence_2020} and is therefore not thermal.

\subsection*{Perspectives on time evolution of the effective spacetime}
An outstanding question in field theory on curved spacetime pertains to the dynamics of entanglement in time evolving systems and in the presence or absence of a horizon.
Up to this point, we have considered continuous-wave excitation of the cavity, in which the time dependence of  $F_\mathrm{p}$ is described by a stationary phase $\ee^{\ci \omega_p t}$ that leads to a stationary effective geometry described by $\phi_0$ on top of which perturbations $\phi_1$ propagate causally.
We now envision a situation in which the temporal dynamics of the system could be controlled as well.

Consider a fluid made of two regions homogeneous in their density and phase.
These quantities can evolve independently in each region, such that the Mach number $M$ in one or both regions changes adiabatically in time.
Through temporal evolution, we keep $M<M_\mathrm{critical}$ in one of the two regions (subcritical flow).
If $M$ grows above $M_\mathrm{critical}$ (supercritical flow) in the other region, a horizon forms.
However, if $M$ grows but never quite exceeds $M_\mathrm{critical}$, no horizon forms (as in Fig.~\ref{fig:ModeStructure} \textbf{(d)}).
In both cases, the temporal evolution of the curvature results in the mixing of positive and negative norm modes of the field and in spontaneous emission of entangled pairs from the vacuum.
If a horizon forms, Hawking radiation will be emitted and might bear imprints of the temporal dynamics~\cite{brout_primer_1995,michel_non-linear_2015,fabbri_rampup_2021}.
However, even if no horizon forms, pairs will still be emitted at the interface between the two regions due to the time-dependence of the fluid profile~\cite{Barcelo:2006uw}.

An instance of such horizonless emission has been predicted to occur in the formation of so-called Exotic Compact Objects (ECOs)~\cite{Barcelo:2010xk}.
These are hypothetical astrophysical objects which are more massive than neutron stars and yet possess no horizon~\cite{Cardoso:2017cqb}.
The formation of ECOs is predicted to lead to a Hawking-like emission if the final state is sufficiently close to a horizon~\cite{Barcelo:2010xk}, hinting at a generalized view of the Hawking effect that relies on quasi-local properties of the spacetime instead of the formation of horizons (which are global properties of spacetimes)~\cite{Barcelo:2010pj,Barcelo:2010xk}.

As time-dependent polariton flows can be engineered by resorting to pulsed instead of CW excitation~\cite{sanvitto_all-optical_2011,wertz_propagation_2012,giorgi_relaxationoscillations_2014}, all the tools are available to experimentally investigate the Hawking effect with and without horizons.

Meanwhile, classical observable properties of ECOs can be arbitrarily close to those of black holes as well~\cite{Cardoso:2016rao}, making their discrimination via state-of-the-art gravitational wave observations~\cite{LIGOScientific:2016aoc,LIGOScientific:2021qlt,NANOGrav:2023gor,LISA:2022yao,LISA:2022kgy,LISA:2022yao,Maggiore:2019uih,Barack:2018yly} and black hole imaging~\cite{EventHorizonTelescope:2019dse,EventHorizonTelescope:2022wkp,EventHorizonTelescope:2020qrl,Perlick:2021aok,Vagnozzi:2022moj} a central challenge in modern astrophysics.

Fast spinning ECOs would be surrounded by an ergosurface and, in the absence of a horizon to dissipate superradiance, are thought to be dynamically unstable~\cite{Brito:2015oca,Cardoso:2014sna,Cunha:2017qtt,Maggio:2017ivp,Maggio:2018ivz,Zhong:2022jke}.
The unfolding of this instability in vortex flows~\cite{giacomelli_ergoregion_2020,patrick_quantum_2022,Torres:2022bto} can be accessed by measuring modes of complex frequency (positive imaginary part).
These modes manifest themselves as plateaus in the dispersion relation, which can be resolved with our methods~\cite{claude_high-resolution_2022},
thereby allowing the study of their imprint on the spectrum of spontaneous emission.

\subsection*{Outlook}
The half-photonic nature of the polariton fluid brings all the tools of quantum optics to the table, which should enable the measurement of entanglement with homodyne detection.
The key to perform such observations is access to the spectrum of excitations, i.e. to the relativistic field theory at play in the experiment.

We have shown that, because of the non-equilibrium fluid dynamics, collective excitations in polaritons are described by a massive scalar field.
The mass gap is directly controlled by setting the driving field intensity and its frequency detuning with respect to the polaritons, while positive and  negative norm mode are excited at higher and lower frequencies than the finite frequency of the driving field.
Our system thus  enables the study of the interplay between field mass and amplification.

We observed the spectrum of collective excitations on the effective spacetime and evidenced the excitation of negative norm modes both in the sub- and supercritical regimes.
Negative norm modes at positive frequencies have negative energy (they are modes which tend to lower the total energy of the system~\cite{Bogolyubov:1947zz,carusottoQuantumFluidsLight2013a}).
Their excitation is the chief signature of the formation of a horizon.

Our methods can be readily generalized to engineer time-dependent and/or vortex flows featuring an ergosurface outside the horizon and to investigate the the dynamics of entanglement in the interplay between the Hawking effect and rotational superradiance~\cite{agullo_entanglement_2023}.
These are but examples of the potential of the fluid of light platform to study open questions in field theory on curved spacetime.

\begin{acknowledgments}
We are thankful to Ferdinand Claude, Iacopo Carusotto and Ivan Agullo for countless and invaluable discussions on polaritons and amplification phenomena.
We thank Tangui Aladjidi for the development of the \textbf{phase-utils} \textit{Python} ~\cite{phaseutils} package that we used in this experiment and Andrew Haky for feedback on the manuscript.
We acknowledge financial support from DIM SIRTEQ project FOLIAGE and from CNRS via a 80prime PhD studentship. QG and AB are members of the Institut Universitaire de France. AD is supported by the NSF grants PHY-
2110273, PHY-1903799, and PHY-2206557, by the RCS
program of Louisiana Boards of Regents through
grant LEQSF(2023-25)-RD-A-04, and by the Hearne Institute for Theoretical Physics
\end{acknowledgments}
\newpage
\begin{appendix}
\begin{widetext}

\section{Collective excitations of a polariton fluid as a massive Klein-Gordon field}\label{App:metric}

In this appendix we derive the analogy between hydrodynamic excitations of a homogeneous polariton fluid and a relativistic massive scalar field. We start with the driven-dissipative GPE \eqref{eq:ddGPE} for the field $\Phi$ describing LPs
\begin{equation}
\ci{{\hbar}}\partial_t\Phi=\lrsq{{{\hbar}}\omega_0-\frac{\hbar^{{2}}}{2m^*}\nabla^2+g|\Phi|^2-\ci\frac{{{\hbar}}\gamma}{2}}\Phi+ F_p.
    \label{eq:dGPE}
\end{equation}
These equations can be written in terms of density phase variables $\Phi=\sqrt{n}\ee^{\ci \theta}$, leading to a continuity equation and an Euler equation for an Eulerian non-barotropic fluid with velocity given by $\pmb{v}=\hbar\vna \theta/m^*$. Steady state configurations can be achieved by pumping with a monochromatic laser. A laser with the form $F_\tp=|F_\tp|\ee^{\ci(\theta_\tp({\bf x})-\omega_\tp t)}$ will drive the polariton field into a steady-state configuration $\Phi=\sqrt{n_0}\ee^{\ci(\theta_\tp({\bf x})-\omega_\tp t)}$, where $n_0$ and $\pmb{v_0}=\hbar\vna\theta_\tp/m^*$ satisfy

\begin{equation}
\lrsq{-\frac{\hbar^2}{2m^*}\nabla^2+\ci\frac{\hbar}{2}(\vna\cdot\pmb{v_0}) +\ci\hbar\pmb{v_0}\cdot\vna- \delta(v_0)+g_{\rm r}n_{\rm r}+g n_0-\ci\frac{\hbar\gamma}{2}}\sqrt{n_0}+|F_\tp|=0\,,
    \label{eq:StatHomogEOSDens}
\end{equation}
with $\hbar \delta(v_0)=\hbar\omega_\tp-\hbar\omega_0-m^* v_0^2/2$ (note that this is just $\delta(k_\tp)$ in a homogeneous configuration where $\pmb{v_0}=\hbar\pmb{k_p}/m^*$) and we have also subtracted the dark reservoir mean energy to the contribution of the detuning . In our notation, a bold letter $\pmb{v}$ is a vector with modulus $v$.

Low-energy (collective) excitations of the polariton fluid around a stationary solution are described by linear perturbations of the polariton field, which can be written in the form $\Phi=(\sqrt{n_0}+\ee^{-\frac{\gamma}{2}t}\phi_1)\ee^{\ci(\theta_\tp-\omega_\tp t)}$. Plugging this into \eqref{eq:dGPE}, and \eqref{eq:StatHomogEOSDens}, we obtain, to linear order in $\phi_1$,
\begin{equation}
    \ci \hbar \lr{\partial_t+\pmb{v_0}\cdot{\vna}}\phi_1=\lrsq{-\frac{\hbar}{2m^*}\nabla^2+\rho-\ci\sigma}\phi_1+g n_0 \phi_1^*,
        \label{eq:DefPertGPE}
\end{equation}
where $$\rho\coloneqq2g n_0- \delta(v_0)+g_{\rm r}n_{\rm r}\qquad\text{and}\qquad\sigma\coloneqq\hbar/2\vna \cdot\pmb{v_0}.$$
Note that here we have assumed that there is no external potential.
This could otherwise be encoded in $\delta(v_0)$.

The above equation and its complex conjugate can be seen as a system of two coupled partial differential equations (PDEs) of first order in time-derivatives and second order in space-derivatives. A way to solve this system is by rewriting it as a single PDE of second order in time and fourth order in space. To that end, we solve $\phi_1^*$ in terms of $\phi_1$ from the above equation to obtain
\begin{equation}
    \phi_1^*=\frac{1}{g n_0}\lrsq{\ci \hbar (\partial_t+\pmb{v_0}\cdot{\vna})+\frac{\hbar}{2m^*}\nabla^2-\rho+\ci\sigma}\phi_1.
\end{equation}
Plugging this into the complex conjugate equation of \eqref{eq:DefPertGPE} we find
\begin{equation}
\begin{split}
    \ci \hbar \lr{\partial_t+\pmb{v_0}\cdot{\vna}}&\frac{1}{g n_0}\lrsq{\ci \hbar \lr{\partial_t+\pmb{v_0}\cdot{\vna}}+\frac{\hbar}{2m^*}\nabla^2-\rho+\ci\sigma}\phi_1=\\
    &\lrsq{\lr{\frac{\hbar}{2m^*}\nabla^2-\rho-\ci\sigma}\frac{1}{g n_0}\lr{\ci \hbar \lr{\partial_t+\pmb{v_0}\cdot{\vna}}+\frac{\hbar}{2m^*}\nabla^2-\rho+\ci\sigma}-g n_0} \phi_1
\end{split}
    \label{eq:SecOrdPhi}
\end{equation}
For homogeneous backgrounds we have $\sigma=0$ and $\delta(v_0)=\delta(k_{\tp})$. Hence, the above equation \eqref{eq:SecOrdPhi} can be simplified to
\begin{equation}
    \lrsq{\lr{\partial_t+\pmb{v_0}\cdot{\vna}}^2+\lr{\frac{\xi^2}{4}\nabla^2-1}\frac{\rho}{m^*}\nabla^2+\frac{\rho^2-g^2n_0^2}{\hbar^2}}\phi_1=0,
\end{equation}
where we have defined the healing length
\begin{equation}
    \xi\coloneqq\frac{\hbar}{\sqrt{m^*\rho}}=\frac{\hbar}{\sqrt{m^*\lr{2gn_0+g_{\rm r}n_{\rm r}-\delta(k_\tp)}}}.
\end{equation}
This length scale is the limit of the hydrodynamical regime. Indeed, for large wavelength perturbations $\lambda\gg\xi$ (low wavenumber $k\ll\xi^{-1}$), we can set  ${\xi^2}\vna^2/4-1\approx-1$. This is called the hydrodynamic approximation, because neglecting this term corresponds to neglecting the effect of the quantum pressure, so that the fluid becomes barotropic in that regime. Indeed,
under this approximation, the above 4th order PDE for $\phi_1$ reduces to a 2nd order PDE of the form
\begin{equation}
    \lrsq{-\lr{\partial_t+\pmb{v_0}\cdot\vna}^2+c_s^2\nabla^2-\frac{m_{\rm det}^2 c_\mathrm{B}^4}{\hbar^2}}\phi_1=0.
    \label{eq:HomogPertWaveEq}
\end{equation}
where 
\begin{equation}\label{eq:csandmdet}
\begin{split}
    &c_\mathrm{B}=\sqrt{\frac{\rho}{m^*}}=\sqrt{\frac{2gn_0- \delta(k_\tp)+g_\mathrm{r}n_\mathrm{r}}{m^*}},
    \\
    &m_{\rm det}=\frac{\sqrt{(\rho^2-g^2n_0^2)}}{c_\mathrm{B}^2}=m^*\frac{\sqrt{(gn_0- \delta(k_\tp)+g_\mathrm{r}n_\mathrm{r})(3gn_0- \delta(k_\tp)+g_\mathrm{r}n_\mathrm{r})}}{(2gn_0- \delta(k_\tp)+g_\mathrm{r}n_\mathrm{r})}.
\end{split}
\end{equation} 
These two quantities have a well defined physical meaning as follows. On the one hand, $m_{\rm det}c_\mathrm{B}^2$ is the energy of $k=0$ modes of $\phi_1$, introducing a mass gap. On the other hand, Eq.~\eqref{eq:HomogPertWaveEq} is a hyperbolic PDE whose characteristic curves in a fluid at rest are given by $|\pmb{x}|=c_\mathrm{B} t$. These characteristics limit the speed at which acoustic excitations $\phi_1$ can propagate information, defining the soundcones in full analogy to light cones. However, due to the mass gap, the speed of propagation of excitations of $\phi_1$ is $v_{\rm g}=d\omega/dk\leq c_\mathrm{B}$. The latter can be seen as the limiting speed of propagation of $k\to\infty$ perturbations. This is also in full analogy to propagation of modes of a massive relativistic scalar field, which travel at subluminal speeds, reaching only the speed of light in the limit of infinite momentum.

Indeed, this analogy can be extended beyond the fluid rest frame, as Eq.~\eqref{eq:HomogPertWaveEq} can  always be written as a Klein-Gordon equation for a massive real scalar field in a curved spacetime, which takes the form
\begin{equation}
    \lrsq{\frac{1}{\sqrt{|\eta|}}\partial_\mu\sqrt{|\eta|}\eta^{\mu\nu}\partial_\nu-\frac{m_{\rm det}^2}{\hbar^2}}\phi_1=0\,
    \label{eq:KGEqApp}
\end{equation}
where the acoustic metric $\eta_{\mu\nu}$ is given by the line element
\begin{equation}
    ds^2=c_\mathrm{B}^{2}\lrsq{\lr{ v_0^2-c_\mathrm{B}^2}\dd t^2-2 \pmb{v_0}\cdot\dd \pmb{x} \dd t + \dd\pmb{x}\cdot\dd\pmb{x}}.
\end{equation}
Thus, we see that, in the hydrodynamic regime, collective excitations of a polariton fluid are described by a massive relativistic scalar field, which becomes massless if the polariton fluid is pumped at the turning point of the bistability loop.

Note that Eq.~\eqref{eq:KGEqApp} with the above form of the acoustic metric is exactly our Eq.~\eqref{eq:massmetric}
This acoustic metric coincides with the one that describes sound waves in an inviscid, irrotational and barotropic fluids according to the Theorem of section 2.3 in~\cite{barcelo_analogue_2011}. However, perturbations in isolated inviscid, irrotational and barotropic fluids are described by a massless Klein-Gordon equation. The breaking of the isolation condition, due to the
driven-dissipative nature of polaritons, modifies the result of that Theorem by introducing a mass term, whose value can be experimentally controlled at will (and even made to vanish) through the detuning $\delta(k_\tp)$.

\section{Speed of sound measurement}\label{App:cs}

\subsection{Dark reservoir contribution}

Selection rules dictate that certain excitonic states do not couple with light. These states form a long-lived excitonic reservoir \cite{stepanovDispersionRelationCollective2019} which impacts the dynamics and contributes an interaction energy $g_\mathrm{r}n_\mathrm{r}$ to be considered to describe the system fully.

For a homogeneous fluid flowing at wavevector $k_\mathrm{p}$ with interaction energy $gn$ and detuning $\delta(k_\mathrm{p})$ we modify the GPE~\eqref{eq:ddGPE} as a set of coupled equations:

\begin{equation}
\ci{{\hbar}}\partial_t\Phi=\lrsq{{{\hbar}}\omega_0-\frac{\hbar^{{2}}}{2m^*}\nabla^2+g|\Phi|^2+g_\mathrm{r}n_\mathrm{r}(r,t)-\ci\frac{{{\hbar}}(\gamma+\gamma_\mathrm{in})}{2}}\Phi+ F_p.
    \label{eq:dGPEannex}
\end{equation}

\begin{equation}
     \partial_t n_\mathrm{r} = -\gamma_\mathrm{r} n_\mathrm{r} + \gamma_\mathrm{in} n.
\label{eq:reservoir_eq}
\end{equation}
where $\gamma_\mathrm{r}$ is the long-lived exciton reservoir decay rate and $\gamma_\mathrm{in}$ the decay rate of polaritons into the reservoir. Taking the steady state of the above equation shows that $n_\mathrm{r}$ and $n$ are proportional through $\gamma_\mathrm{r} n_\mathrm{r} =  \gamma_\mathrm{in}n$.
This energy renormalization modifies the spectrum of collective excitations as well, see Eqs.~\eqref{eq:ffbogo} and~\eqref{eq:lfbogo} in the main text.

We measure a polariton decay rate $\hbar\gamma \approx \SI{80}{\micro \electronvolt}$.
Meanwhile, the reservoir contribution $\beta$ depends on the exciton-photon detuning $\Delta E_{X-\gamma} = E_X - E_\gamma$.
We use the same sample as in~\cite{claude_spectrum_2023} and ensure the same $\Delta E_{X-\gamma}$. We therefore use
\begin{equation}
    g_\mathrm{r}n_\mathrm{r} \approx 1.84gn
    \label{eq:reservoir_prop}
\end{equation}
in Eq.~\eqref{eq:lfbogo} where experimental data are fitted to extract $gn$.

\begin{figure}[h!]
    \centering
\includegraphics[width=0.5\textwidth]{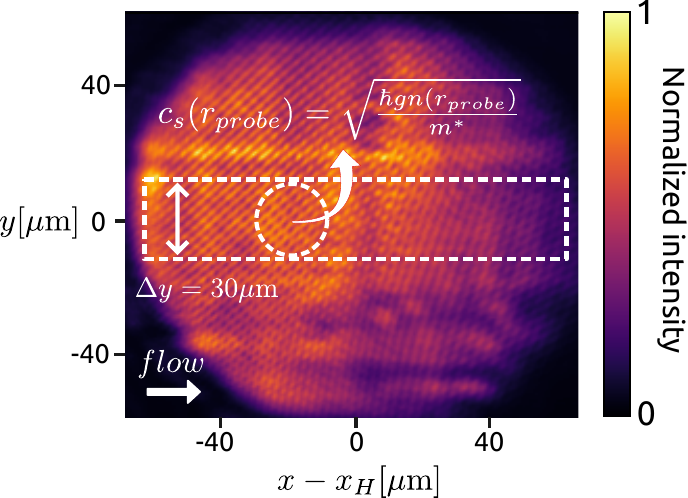}
    \caption{
    Real space intensity measured for a typical fluid that corresponds to the spectrum measured in Fig~\ref{fig:ModeStructure} \textbf{(d)}  (blue). The white dashed rectangle shows the region of interest at which all the measurements were made. The white dashed circle shows the position of the probe laser used to measure the spectrum [Fig~\ref{fig:ModeStructure} \textbf{(d)} ] from which $c_\mathrm{s}(r_\mathrm{probe})$ is extracted.
    }
    \label{fig:S1}
\end{figure}

\subsection{Effects of the spatial extension of the fluid}
Since the created fluids have a certain spatial extension $\Delta y \approx \Delta x \approx \SI{100}{\micro\meter}$ some of the parameters of both the optical microcavity and the exciting field may change in space. In this picture, the way we infer intracavity observables from the photonic signal escaping the sample may change from one point of the fluid to another. In this appendix, we show that, given that the field outcoming the sample is directly proportional to the intracavity field, we can locally measure the speed of sound $c_\mathrm{s}(r)$.

The lower polariton field can be expressed through its excitonic and photonic parts as
\begin{equation}
    \hat{\phi}= C_k\hat{a}_k+X_k\hat{b}_k,
\end{equation}
where $\hat{a_k}$ and $\hat{b_k}$ are respectively the microcavity photon and the exciton annihilation operators. The Hopfield coefficients $C^2_k$ and $X^2_k$ give respectively the proportion of photons and excitons in a given polariton mixed state through 
\begin{equation}
    C_k^2 = \frac{\sqrt{\Delta E_{X-\gamma}(k)^2+\hbar^2\Omega_R^2}+\Delta E_{X-\gamma}(k)}{2\sqrt{\Delta E_{X-\gamma}(k)^2+\hbar^2\Omega_R^2}}  \qquad
     X_k^2 = \frac{\sqrt{\Delta E_{X-\gamma}(k)^2+\hbar^2\Omega_R^2}-\Delta E_{X-\gamma}(k)}{2\sqrt{\Delta E_{X-\gamma}(k)^2+\hbar^2\Omega_R^2}},   
\label{eq:Hopfield}
\end{equation}
where $\Delta E_{X-\gamma}(k) = E_X(k)-E_\gamma(k)$ is the exciton-photon detuning at wavevector $\mathbf{k}$ and $\hbar\Omega_R$ is the vacuum Rabi splitting.

In the experiment, we collect the polariton photonic component $C_k$ that carries all the fluid information.As shown in Eq.~\eqref{eq:Hopfield}, as soon as the effective detuning $\Delta E_{X-\gamma}(k)$ changes, the photonic fraction of the intracavity field is modified. The wavevector changes continuously from $k(x\ll x_\mathrm{H})=k_\mathrm{u}$ to $k(x\gg x_\mathrm{H})=k_\mathrm{d}$, from the sub- to the supercritical asymptotic regions.
As a result, the effective detuning changes locally.
For a typical fluid [Fig~\ref{fig:ModeStructure} \textbf{(d)} (green)] we have $k_\mathrm{u}=\SI{0.11}{\per\micro\meter}$ and $k_\mathrm{d}=\SI{0.63}{\per\micro\meter}$, giving $C^2_{k_\mathrm{u}}=0.52$ in the upstream region and $C^2_{k_\mathrm{d}}=0.51$ in the downstream region. We can thus consider that the coefficient relating the intracavity field amplitude to its photonic component doesn't change in space on the whole fluid and is $C^2= 0.51 $. 

The energy landscape seen by the fluid is also changed by the presence of a wedge between the mirrors, which is due to the growth method, resulting in a spatial change in the photon resonance in a given direction $\mathbf{\vec{u}}$ scaling as $w =\SI{0.04}{\micro\electronvolt\per\micro\meter}$. By orienting the sample in such way that $\mathbf{\vec{u}}$ is orthogonal to the $x$-axis (the propagation direction of the fluid) $\mathbf{\vec{u}} \cdot \mathbf{{\vec{x}}}=0$, only the spatial extension $\Delta y $ of the fluid in the y direction feels the wedge. Typically, for $\Delta y = \SI{30}{\micro\meter}$, $\Delta E_y = \SI{1.2}{\micro\electronvolt}$. This results in a negligible change in the Hopfield coefficients and thereby the effective 1D description of the fluid is fully justified.
In practice, we record the output intensity map Fig~\ref{fig:S1}, which is proportional to the intracavity density  
\begin{equation}
I_\mathrm{out}(r) \approx C^2 |\phi(r)|^2
\label{eq:Iout}
\end{equation}
We then center the probe at a given location $r_\mathrm{probe}$ (dashed line in Fig.~\ref{fig:S1}) and perform a measurement of the Bogoliubov spectrum from which we extract the interaction energy at  $gn(r=r_\mathrm{probe})$ giving the speed of sound $c_\mathrm{s}(r_{probe})=\sqrt{\hbar( g n(r_\mathrm{probe})-\delta(k_\mathrm{p})+g_\mathrm{r}n_\mathrm{r}) /m^*}$. From \eqref{eq:Iout} we calculate the local speed of sound at any point of the fluid as
\begin{equation}
    c_\mathrm{s}(r)=c_\mathrm{s}(r_\mathrm{probe})\sqrt{\frac{I_\mathrm{out}(r)}{I_\mathrm{out}(r_\mathrm{probe})}}.
\end{equation}

\section{Critical velocity with near-resonance excitation}\label{app:vctirical}
When the cavity is pumped near-resonance, the detuning between the pump frequency $\omega_\mathrm{p}$ and the frequency of the bare polaritons, $\delta(k_\mathrm{p})$, controls the optical bistability cycle~\cite{baas_bista_2004}.

At the so-called turning point of the optical bistability cycle, $gn=\delta(k_\mathrm{p})-g_\mathrm{r}n_\mathrm{r}$ and the dispersion relation in the laboratory frame~\eqref{eq:lfbogo} simplifies to
\begin{equation}
    \label{eq:TPbogo}
        \omega_\mathrm{B}^{\pm}(\delta k)= v_0(x) \delta k(x)\,\pm\sqrt{\frac{\hbar \delta k(x)^2}{2m^*}\left(\frac{\hbar \delta k(x)^2}{2m^*}+2gn(x)\right)}-\frac{\ci\gamma}{2}    
\end{equation}
with $\delta k(x)=k-k_\mathrm{p}(x)$.
In this case, the critical velocity to excite negative energy waves simply coincides with the speed of sound $c_\mathrm{s}$.

However, if $gn>\delta(k_\mathrm{p})-g_\mathrm{r}n_\mathrm{r}$, a mass gap opens in the dispersion, with $m_{\rm det}=m^*\sqrt{(gn_0- \delta(k_\tp)+g_\mathrm{r}n_\mathrm{r})(3gn_0- \delta(k_\tp)+g_\mathrm{r}n_\mathrm{r})}/(2gn_0- \delta(k_\tp)+g_\mathrm{r}n_\mathrm{r})$.
We rewrite the dispersion in the laboratory frame~\eqref{eq:lfbogo} in a form that allows to easily identify the effect of the mass on the spectrum,
\begin{align}\label{eq:lfbogom}
    \omega^\pm_\mathrm{B}(\delta k)=&v_0(x) \delta k(x)-\ci\frac{\gamma}{2}\nonumber\\&\pm\sqrt{\left(\frac{\hbar\delta k(x)^2}{2 m^{*}}\right)^2+\frac{\hbar \delta k(x)^2}{m^*}\lr{2g n_0 - \delta(k_\mathrm{p})+g_\mathrm{r}n_\mathrm{r}}+\lr{g n_0 - \delta(k_\mathrm{p})+g_\mathrm{r}n_\mathrm{r}}\lr{3 g n_0 - \delta(k_\mathrm{p})+g_\mathrm{r}n_\mathrm{r}}}\nonumber\\
    =&v_0(x) \delta k(x)-\ci\frac{\gamma}{2}\,\pm\sqrt{\left(\frac{\hbar\delta k(x)^2}{2 m^{*}}\right)^2+c_\mathrm{B}^2\left(\delta k(x)^2+m_\mathrm{det}\right)}
\end{align}
where $c_\mathrm{B}$ defines the sound cones, see Eq.~\eqref{eq:csandmdet}.

From Eq.~\eqref{eq:lfbogom}, we see that the field mass $m_\mathrm{det}$ increases $\abs{\mathrm{Re}(\omega_\mathrm{B}^{\prime\pm})}$ at all $k$, meaning that the critical velocity to have a Doppler effect large enough to excite negative energy  waves is larger than $c_\mathrm{s}$.

The value of the critical velocity at which negative energy waves become available at positive frequency can be computed as follows.
At vanishing flow velocity $k_p=0$ and for $gn>\delta(0)-g_\mathrm{r}n_\mathrm{r}$, the dispersion $\omega^\pm$ branches are symmetric parabolas around the vertical axis.
The negative/positive branches having respectively a maximum/minimum at $\omega_B^\pm(0)=\pm m_{\rm det}c_\mathrm{B}^2/\hbar$ due to the mass gap.

As $k_p$ increases, the negative norm branch is displaced to larger $k$ and pulled to higher $\omega$, while the positive norm branch is displaced to lower $k$ and pulled to lower $\omega$.
There is a critical value of $k^{\rm crit}_p$ at which the extrema of both branches touch the horizontal axis at some $\delta k=\delta k_0\neq0$, namely at which $\omega_{\rm B}^\pm(\delta k_0)=0$. 

For $k_p>k^{\rm crit}_p$, there will be negative norm modes available at positive frequencies, and there will be two values $\delta k_{1,2}$ at which each dispersion branch intersects the horizontal axes. These $\delta k_{1,2}$ depend on the value of $k_p$. 
The critical wavenumber $k_p^{\rm crit}$, and correspondingly the critical velocity $v_{\rm crit}=\hbar k_p^{\rm crit}/m^*$, is obtained when the two intersection points merge, so that $\delta k_1=\delta k_2=\delta k_0$.
The critical velocity can generally be found by solving the implicit equation
\begin{equation}
    v_{\rm crit}=c_\mathrm{B}\sqrt{1+\frac{m_{\rm det}(v_{\rm crit})}{m^*}}\,,
\end{equation}
for $v_{\rm crit}$. Note that $m_{\rm det}$ generally depends on the velocity of the fluid through $k_p$, so that the critical velocity is the value of the flow velocity that solves the above equation once all the dependence of $m_{\rm det}$ in terms of the flow velocity is explicitly written. This dependence makes an explicit formula for $v_{\rm crit}$ rather involved, and not particularly elucidating.

Because of the increased critical velocity, the Mach number that must be exceeded to have a supercritical dispersion relation is $M_\mathrm{critical}=v_\mathrm{critical}/c_\mathrm{B}=\sqrt{1+m_\mathrm{det}/m^*}>1$.
We take the long-lived excitonic reservoir contribution to be $\beta=1.84$~\cite{claude_spectrum_2023} so, for typical parameter values in the experiment, eg the configurations of Fig.~\ref{fig:ModeStructure} ($k_\mathrm{p}=\SI{0.55}{\per\micro\meter}$, $\delta(k_\mathrm{p})=\SI{27}{\micro\electronvolt}$, $\hbar gn=\SI{10}{\micro\electronvolt}$), we have $M_\mathrm{critical}=1.34$.

\end{widetext}
\end{appendix}
\end{document}